# Rheological Investigation of The Network Structure in Mixed Gels of Kappa and Iota Carrageenan

Tulika Bhattacharyya, [a)] Chandra S Palla, [b)] Dattatraya H. Dethe, [(c)] and Yogesh M. Joshi* [a)]

[a)] Department of Chemical Engineering, Indian Institute of Technology Kanpur, Kanpur, Uttar Pradesh 208016, India

[b)] S. C. Johnson, Racine, Wisconsin, 53403, United States

[c)] Department of Chemistry, Indian Institute of Technology Kanpur, Kanpur, Uttar Pradesh 208016, India

*Corresponding Author, email: joshi@iitk.ac.in

**Abstract:**

Carrageenans comprise linear sulfated high molecular weight polysaccharides obtained from seaweeds and are routinely used in food and home/personal care industries. Various kinds of carrageenans differ from others based on the ester sulfate group location on the polysaccharide chains. Pure and mixed systems of Kappa Carrageenan and Iota Carrageenan undergo a three-dimensional gel network structure formation or dissociation with a change in temperature. During the sol-gel and gel-sol transitions, the Carrageenan systems pass through a unique critical gel state, where dynamic moduli are scale-invariant owing to the self-similar structure of the three-dimensional network. In this work, we obtain the critical gel state associated with pure and mixed systems of Kappa and Iota Carrageenan during cooling and heating by exploring the material behavior for a range of frequencies. Interestingly, on the one hand, the mixed gels show a higher critical sol-gel transition temperature compared to the pure systems at equal individual concentrations. On the other hand, the low temperature moduli of mixed gels are closer to that of Kappa Carrageenan when the concentration of the same is more than half in the mixture. The rheological measurements demonstrate that the Kappa Carrageenan strongly affects the nature of aggregation of double helices of Iota Carrageenan, but Iota Carrageenan does not have a significant influence on that of Kappa Carrageenan. These results suggest an associative, interactive network formation between Kappa and Iota Carrageenan in the mixture, such that the gel behavior is predominantly influenced by Kappa Carrageenan.

**Graphical Abstract:**

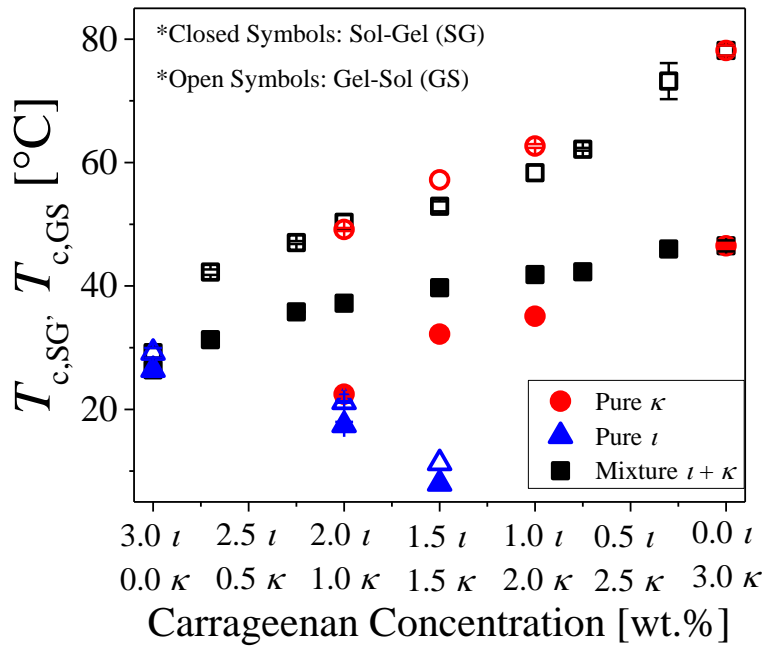

**Graphical Abstract Figure:** The figure represents the critical sol-gel and gel-sol transition temperature associated with different concentrations of the pure and mixed systems of Kappa and Iota Carrageenan. The extent of thermoreversibility in the mixed gel decreases with increase in the Kappa fraction in the mixed system.

1. **Introduction:**

   Carrageenan, a derivative of red seaweed algae, is a family of naturally occurring linear sulphated polysaccharides. The members of the carrageenan family are categorized based on the positions and number of sulphate groups in the disaccharide repeating units (Campo, Kawano, Silva, & Carvalho, 2009; Estevez, Ciancia, & Cerezo, 2000; Imeson, 2009; van de Velde & Rollema, 2006). Owing to network formation through temperature control, carrageenans are routinely used in food and beverages, biomedical and biotechnological industries and pharmaceutical products as gelling agents, viscosity enhancers and stabilizers (Falcó, Randazzo, Sánchez, López-Rubio, & Fabra, 2019; Hansen, 1993; Langendorff, et al., 2000; Li, Ni, Shao, & Mao, 2014; Pacheco-Quito & Ruiz-Caro, 2020; Rhein-Knudsen & Meyer, 2021). The Kappa Carrageenan ($\kappa-C$) and Iota Carrageenan ($\iota-C$), being the two most commercially used gel-forming agents of this family, have received significant attention from the scientific community. As shown in the structural schematic Figure 1, the $\kappa-C$ consists of repeating units of $\beta(1-3)$-D-galactose-4-sulfate and $\alpha(1-4)$-3,6 anhydro-D-galactose. The $\iota-C$ has the similar (1-3) linked galactose unit but consists of an additional sulphate group in the (1-4) linked galactose unit. It has been reported that upon cooling the aqueous solutions of pure $\kappa-C$ and $\iota-C$, the random coils of the same undergo helix formation. This causes aggregation of double helices, leading to formation of a three dimensional gel network. Strikingly, while the gels of $\kappa-C$ are rigid and brittle with strong thermal irreversibility, $\iota-C$ leads to formation of softer, thermoreversible gels (Bui, Nguyen, Nicolai, & Renou, 2019a; Geonzon, et al., 2023; Hossain, Miyanaga, Maeda, & Nemoto, 2001; Liu, Chan, & Li, 2015; Parker, Brigand, Miniou, Trespoey, & Vallée, 1993; Thrimawithana, Young, Dunstan, & Alany, 2010). This distinct nature of the $\kappa-C$ and $\iota-C$ gels arise from the inherent difference in charge density in the two polysaccharides. Interestingly, the chains of these two polysaccharides coexist together naturally and often disperse as short or long sequences in chains of each other. In order to obtain gels with intermediate rigidity and in order to have greater flexibility in designing gel characteristics, $\kappa-C$ and $\iota-C$ are often combined to form mixed gels. Accordingly, it is necessary to understand the gelation mechanism and microstructure of the mixed gels of $\kappa-C$ and $\iota-C$, which have very similar repeating units but different charge densities (Bui, Nguyen, Renou, & Nicolai, 2019; Geonzon, Descallar, Du, Bacabac, & Matsukawa, 2020).

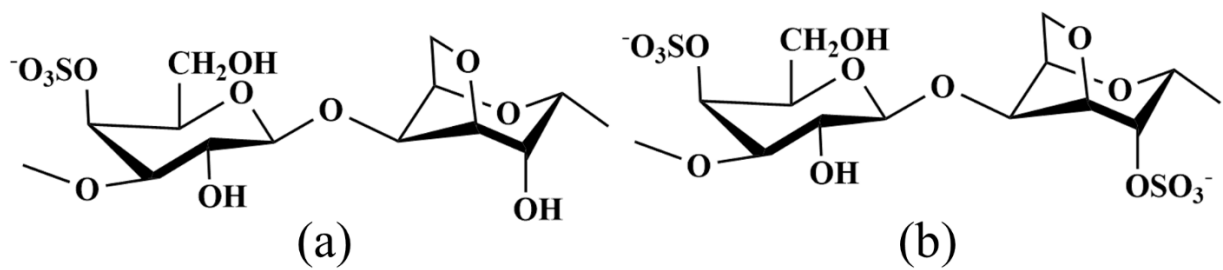

**Figure 1.** Chemical structure of the (a) Kappa Carrageenan ($\kappa-C$) and (b) Iota Carrageenan ($\iota-C$) monomer. The (1-3) linked galactose unit is similar in both $\kappa-C$ and $\iota-C$. The (1-4) linked galactose unit comprises an additional sulphate group in $\iota-C$.

In a rheological experiment, starting from high temperature solution, the elastic modulus of pure and mixed systems of $\kappa-C$ and $\iota-C$ increase as the gels consolidate with a decrease in temperature. The temperature at which an increase in the elastic modulus of the sol shows an inflation has been considered as the gelation temperature by many researchers (Bui, Nguyen, Renou, et al., 2019; Du, Brenner, Xie, & Matsukawa, 2016; Geonzon, Bacabac, & Matsukawa, 2019a, 2019b; Parker, et al., 1993). Very interestingly, the mixed gels demonstrate a distinct two-step increase (two inflation points) in elastic moduli, which is otherwise absent in pure gels (Brenner, Tuvikene, Parker, Matsukawa, & Nishinari, 2014; Bui, Nguyen, et al., 2019a; Bui, Nguyen, Renou, et al., 2019; Du, Brenner, et al., 2016; Geonzon, et al., 2019a; Parker, et al., 1993; Piculell, Nilsson, & Muhrbeck, 1992). This behavior was first noted in $\iota-C$ gels with $\kappa-C$ impurity, in presence of sodium and potassium cations (Parker, et al., 1993; Piculell, et al., 1992). Similar two-step change in properties during gelation have also been reported in thermograms of Dynamic Scanning Calorimetry (DSC) (Brenner, et al., 2014; Ridout, Garza, Brownsey, & Morris, 1996) and optical rotation measurements as a function of temperature (Piculell, Nilsson, & Ström, 1989; Rochas, Rinaudo, & Landry, 1989). In the literature on mixed gels, the temperatures corresponding to the two-step changes in properties have been considered as the individual gelation temperature of each type of carrageenan (Bui, Nguyen, Renou, et al., 2019; Du, Brenner, et al., 2016; Du, Lu, Geonzon, Xie, & Matsukawa, 2016; Geonzon, et al., 2019a; Parker, et al., 1993). Accordingly, the $\kappa-C$ and $\iota-C$ have been proposed to develop separate network structures in the mixed system (Parker, et al., 1993). However, there is a major inconsistency in the literature while comparing the gelation temperature for the mixed system with pure components utilizing the two-step change. On the one hand, some researchers correlate the two-step changes in mixed systems with the independent gelation temperatures of pure components but at the same total carrageenan concentration (Du, Brenner, et al., 2016; Geonzon, et al., 2019a). Others, however, find that the temperatures corresponding to two-step changes are associated with the concentrations of the individual components in the mixed system. (Bui, Nguyen, Renou, et al., 2019; Parker, et al., 1993).

In the literature, the network structure in mixed gels has frequently been described as either independently interpenetrated or microphase separated. Hu *et al.* (Hu, Du, & Matsukawa, 2016) demonstrated the formation of a homogenous interpenetrating network through NMR studies. According to them, if at all a microphase separated network gets formed, it should be below the lengthscale of 400 nm. Parker *et al.* (Parker, et al., 1993) and Bui *et al.* (Bui, Nguyen, Renou, et al., 2019) proposed that a mixed gel with independent interpenetrated networks must have a modulus that is of the order of the sum of the moduli of individual components. However, since this is not observed for the mixed gels in the presence of added cations, Parker *et al.* (Parker, et al., 1993) and Bui *et al.* (Bui, Nguyen, Renou, et al., 2019) dismissed the possibility of an independent, interconnected network. In the case of microphase separation in mixed gels, it has been proposed that each carrageenan network becomes dense in the presence of the other, leading to a higher modulus than a simple arithmetic sum of the modulus of pure components (Brenner, et al., 2014; Bui, Nguyen, Renou, et al., 2019). In a typical procedure, gel modulus obtained in the low temperature regime at small deformations has been interpreted in terms of the polymer blending laws proposed by Davies (Davies, 1971):

$$G_m^\chi = \varphi_1 G_{p,1}^\chi + \varphi_2 G_{p,2}^\chi \qquad (1)$$

where, $G_{p,1}$ and $G_{p,2}$ are the moduli of components 1 and 2 having phase volume fractions $\varphi_1$ and $\varphi_2$ respectively, while $G_m$ is the modulus of the mixed gel. The magnitude of the exponent $\chi$ determines the type of microphase separation in the network, such that $\chi = \pm 1$ suggests a single continuous phase with the other as filler, while $\chi = 1/5$ suggests a bicontinuous network (Morris, 2009). Rheological measurements of Brenner and coworkers (Brenner, et al., 2014; Du, Brenner, et al., 2016) exhibit $\chi \approx 1/5$, suggesting bicontinuous microphase separated network in mixed gels. Furthermore, particle tracking measurements by Geonzon *et al.* (Geonzon, et al., 2019a) demonstrate structural inhomogeneity, indicating phase separated $\kappa-C$ and $\iota-C$ rich domains of size larger than 100 nm.

The gelation in carrageenan solutions has also been studied using microscopy (Bui, Nguyen, Renou, et al., 2019; Lundin, Odic, Foster, & Norton) and turbidity measurements (Brenner, et al., 2014; Bui, Nguyen, Renou, et al., 2019). Bui *et al.* (Bui, Nguyen, Renou, et al., 2019) performed microscopy and turbidity analysis for pure and mixed gels of $\kappa-C$ and $\iota-C$ and ruled out microphase separation in mixed gels. They reported that the $\iota-C$ network to be homogeneous up to 100 nm in both pure and mixed systems. On the other hand, the pure $\kappa-C$ network is heterogeneous for a few micrometres, which decreases in the mixed system. They observe mixed gel to have lesser turbidity compared to the pure $\kappa-C$. While they observe $\chi \approx 1/5$, they consider it to be misleading. According to their measurements, microphase separation is only possible at lengthscale lesser than 100 nm. However, given that the chains of carrageenan polysaccharide spans about 400 nm, it is difficult to maintain a separated network at this lengthscale. In the literature, many groups have carefully studied the gelation of $\kappa-C$ and $\iota-C$ systems. In Table 1, we present a summary of the type of network structure formation in mixed gels.

**Table: 1.** List of studies on investigation of the type of network structure formation in mixed gels of Kappa and Iota Carrageenan

| Literature | Concentration | Methods | Suggested nature of interactions |
|---|---|---|---|
| Picullel *et al.* (Piculell, et al., 1989) | Car: 1 wt.% rubidium and sodium carrageenan mixture<br>Salt: 100 mM RbCl<br>Purity: Purified | NMR spectroscopy; Optical Rotation | Separated network |
| Picullel *et al.* (Piculell, et al., 1992) | Car: Native and pure $\iota-C$ 1% w/v and 2.5% w/v<br>Salt: 100 mM and 250 mM NaCl or KCl<br>Purity: Purified | Dynamic Rheology | Bicontinuous phase separated network |
| Parker *et al.* (Parker, et al., 1993) | Car: $\iota-C$: 0.5 – 2 wt.%; In 1.5 wt.% $\iota-C$, $\kappa-C$ replaced till 10%<br>Salt: 200 mM KCl +NaCl<br>Purity: Variable | Dynamic Rheology, Cutting force measurements | Phase separated interpenetrating network |
| Ridout *et al.* (Ridout, et al., 1996) | Car: Maximum concentration 2% w/v<br>Salt: 80 mM KCl<br>Purity: $\kappa-C$: 95%; $\iota-C$: 95% | DSC | Phase separated or Interpenetrated networks |
| Lundin *et al.* (Lundin, et al.) | Car: 0-1.5 wt.%<br>Salt: 400 mM NaCl, 50 mM KCl<br>Purity: $\kappa-C$: 98%; $\iota-C$: 98% | Turbidity measurements, DSC, Dynamic Rheology | Bicontinuous phase separated network |
| Brenner *et al.* (Brenner, et al., 2014) | Car: Maximum concentration 1.6 wt.%<br>Salt: 150 mM KCl<br>Purity: $\kappa-C$: 80.5%; $\iota-C$: 98.3% | Rotational and extensional Rheology, Texture analysis, DSC, Turbidity measurements | Bicontinuous phase separated network at large $\iota-C$: $\kappa-C$ ratio |
| Hu *et al.* (Hu, et al., 2016) | Car: Maximum 2wt.%, mixture containing (1:1)<br>Salt: 30 mM KCl<br>Purity: - | Diffusion and proton spin-spin relaxation measurements, Dynamic Rheology, DSC | Homogenous interpenetrating network or microphase separation below 450 nm |

| Literature | Concentration | Methods | Suggested nature of interactions |
| --- | --- | --- | --- |
| Du *et al.* (Du, Brenner, et al., 2016) | Car: 0.5 to 3 wt.% <br> Salt: 10 mM KCl <br> Purity: $\kappa-$C: 97.2%; $\iota-$C: 98.3% | Dynamic Rheology; DSC; Water, and ion migration | Phase separation |
| Du *et al.* (Du, Lu, et al., 2016) | Car: Pure: 1 wt.%, 2 wt.%; Mixed: 1+1 wt.% <br> Salt: 10 mM and 17 mM KCl <br> Purity: $\kappa-$C: 97.2%; $\iota-$C:98.3% | Dynamic Rheology; Multiple particle tracking measurements | Phase separation |
| Geonzon *et al.* (Geonzon, et al., 2019a) | Car: 1.5 wt.%, <br> Salt: 10 mM KCl <br> Purity: - | Particle tracking measurements, Dynamic Rheology | Phase separation |
| Wurm *et al.* (Wurm, Nussbaumer, Pham, & Bechtold, 2019) | Car: Total: 1wt.% <br> Salt: 15 mM $Ca^{2+}$; 50 mM $K^+$; 20mM $Na^+$ <br> Purity: $\kappa-$C: 87%; $\iota-$C: 90% | Dynamic Rheology | Bicontinuous phase separated network |
| Bui *et al.* (Bui, Nguyen, Renou, et al., 2019) | Car: Pure: 1wt.%, 2 wt.%; Mixed: 1+1 wt.%, <br> Salt: 0-100 mM $CaCl_2$; 50 mM KCl <br> Purity: $\kappa-$C: 95%; $\iota-$C: 95% | Confocal Laser Scanning Microscopy, Dynamic Rheology, Turbidity Measurements | Coaggregation of double helices formed by $\iota-$C and $\kappa-$C, Interpenetration with each other |
| This work (Bhattacharyya *et al.*) | Car: Pure: 1 to 3 wt.%, Mixed: 3 wt.%; <br> No added salt; <br> Purity: Purified | Dynamic Rheology Measurements | Associative network structure either through co-double helices formation or coaggregation of double helices formed by $\iota-$C and $\kappa-$C |

The present work utilizes rheology to analyze gelation in Carrageenan solutions. In the earlier rheological studies on mixed gels of $\kappa-$C and $\iota-$C, the frequency dependence of the viscoelastic properties, particularly the dynamic moduli and loss tangent, in the sol and gel states were not explored. As the aqueous solutions of carrageenan undergo sol-gel and gel-sol transition with change in temperature respectively through formation or dissociation of three dimensional network structures, the systems, in principle, should pass through a unique critical gel state (Bui, Nguyen, Nicolai, & Renou, 2019b; Geonzon, et al., 2020; Geonzon, et al., 2023; Liu, et al., 2015; Liu, Huang, & Li, 2016). The critical gel is a distinct state, at which the material forms the weakest space spanning three dimensional network structure (Horst Henning Winter & Mours, 1997). Owing to the fractal like network structure of the critical gel state, the relaxation modes associated with this state are exceptionally coupled in nature, leading to a scale invariant behavior of material properties. Rheometry is a powerful probe to accurately detect the critical gel state during the sol-gel and gel-sol transition. At the critical gel state, the magnitude of the loss tangent, which is the ratio of viscous to elastic moduli, becomes independent of applied angular frequency (Chambon & Winter, 1985; H. Henning Winter & Chambon, 1986; Horst Henning Winter, et al., 1997). Furthermore, the fractal dimension of the self-similar network structure can be computed from the magnitude of loss tangent (Muthukumar, 1989). In the literature, precise knowledge of fractal dimension and viscoelastic properties at the critical gel state has been routinely utilized to understand the microstructure and the gelation mechanism in various systems undergoing physical crosslinking as a function of temperature (Bhattacharyya, Suman, & Joshi, 2023; Djabourov, Leblond, & Papon, 1988; Joshi, Suman, & Joshi, 2020; Liu, et al., 2015; Liu, Huang, et al., 2016; Suman & Joshi, 2020; Suman, Sourav, & Joshi, 2021) as well as time (Cho & Heuzey, 2008; Jatav & Joshi, 2017; Negi, Redmon, Ramakrishnan, & Osuji, 2014; Suman & Joshi, 2019, 2020; Suman, Mittal, & Joshi, 2020). The present work aims to extend this approach to investigate the microstructure during the gelation of the mixed Kappa and Iota Carrageenan systems. The primary objective of this work is to investigate the frequency-independent gelation transition temperatures of mixed systems and compare these with that of pure systems. The measurement of frequency-independent critical gel state also addresses the disagreement regarding the relationship between the temperatures at which mixed and pure systems undergo gelation. During the sol-gel and gel-sol transition, we explore the evolution of the linear viscoelastic properties. Subsequently, we show that analysis of material properties at the critical gel state and in the low temperature regime together provides valuable information about the type of network structure formation in mixed gels.

## 2. Materials and Methods:

The materials $\kappa-C$ and $\iota-C$ have been procured from market and used without any further purification. The $\kappa-C$ sample contains certain amount of externally added NaCl, as declared by the manufacturer. We perform $^1$H NMR measurements of both the samples. The samples for NMR measurements were prepared by addition of approximately 15 mg of the procured carrageenan sample in 0.5 ml of D$_2$O. The NMR measurements are carried out in a 500 MHz spectrometer at a temperature of 70°C. Approximately 120 scans are required for recording of spectra for each sample. The $^1$H NMR spectra for both the Carrageenan samples are reported in Figure 2. The $^1$H NMR spectra of $\kappa-C$ shows fourteen peaks at 5.09, 4.82, 4.63, 4.61, 4.51, 4.37, 4.25, 4.12, 4.04, 3.96, 3.79, 3.59, 3.41 and 3.34 ppm. The $^1$H NMR spectra of $\iota-C$ shows twelve peaks at 5.46, 5.29, 4.89, 4.83, 4.65, 4.36, 4.24, 4.21, 3.97, 3.81, 3.61 and 3.34. It is to be noted that some additional peaks arise due to impurities in the system. However, the NMR spectra matches very well with previous literature on NMR spectra of Carrageenan (Campo, et al., 2009; van de Velde, et al., 2006). This suggests that the samples used in the present work are of analytical grade.

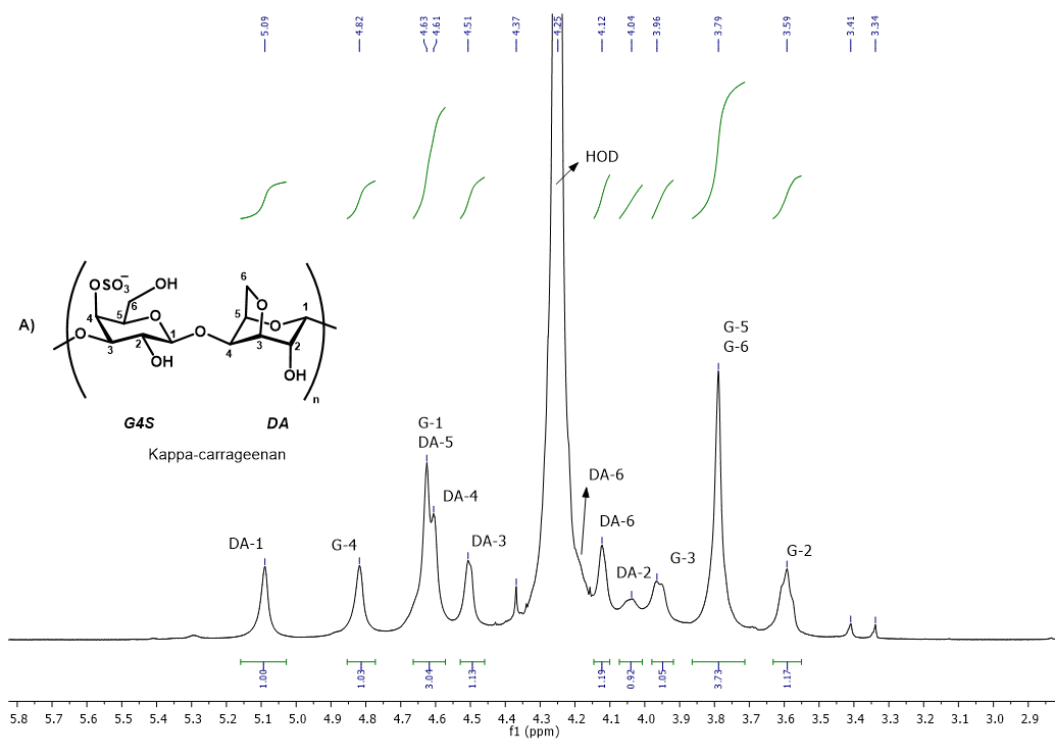

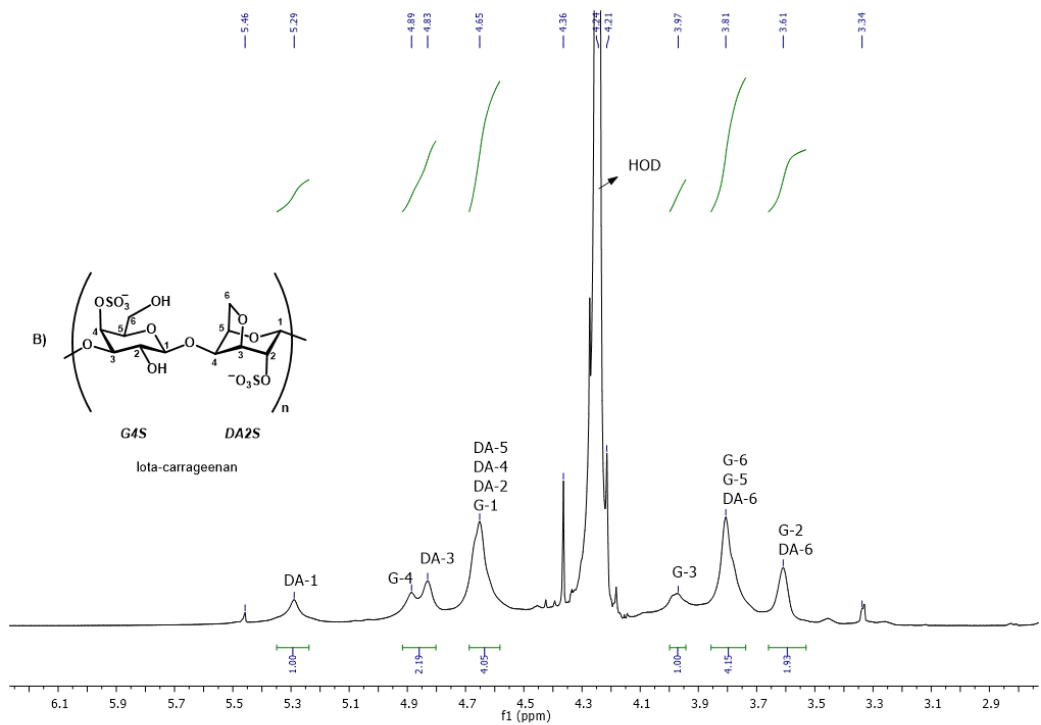

**Figure 2.** $^1$H NMR Spectra for (a) Kappa Carrageenan ($\kappa-C$) and (b) Iota Carrageenan ($\iota-C$).

We obtain molecular weight of both the kinds of Carrageenan using intrinsic viscosity measurement method. The detailed procedure of obtaining molecular weight of Carrageenan from intrinsic viscosity has been discussed by Berth *et al.* (Berth, Vukovic, & Lechner, 2008). Accordingly, we prepare the $\kappa-C$ and $\iota-C$ solution in 0.1 M NaNO$_3$ and obtain the intrinsic viscosity using Ubbelohde viscometer immediately after preparing the solution. At 25°C the intrinsic viscosities of $\kappa-C$ and $\iota-C$ are respectively given by: $[\eta]_{\kappa-C} = 550$ ml/g and $[\eta]_{\iota-C} = 1310$ ml/g. Using the Mark-Houwink plot given by Berth *et al.* (Berth, et al., 2008) the corresponding molar mass of $\kappa-C$ and $\iota-C$ respectively comes out to be: $M_{w,\kappa-C} \approx 2.4 \times 10^5$ g/mol and $M_{w,\iota-C} \approx 8.8 \times 10^5$ g/mol. The ion concentrations of the two samples are measured by preparing 10 ppm aqueous solutions with 2% of added HNO$_3$. The ion concentration measurements are performed in the Agilent ICPMS 7900. The ion concentrations in each of the $\kappa-C$ and $\iota-C$ samples are listed in Table 2. It should be noted that, in the present work, we do not perform mineralization prior to the measurement of the metal ions concentration through ICPMS. Consequently, the data that has been reported only accounts for the concentration of metal ions that dissociate upon dissolution of the sample in the aqueous media. While non-mineralization may have some effect on the total ion concentration, the measurements are repeatable on following the same procedure. Furthermore, in principle, $\iota-C$ should have higher ion concentration compared to $\kappa-C$. However, in the present case we find $\kappa-C$ to have a higher ion concentration owing to externally added sodium chloride by the manufacturer.

**Table 2.** Ion concentrations for $\kappa-C$ and $\iota-C$ measured through Agilent ICPMS 7900.

| Type of ion | $\kappa-C$ | $\iota-C$ |
|---|---|---|
| $Na^+$ ($M_w = 23$ g/mol) | 3.81 % | 2.28 % |
| $Mg^{2+}$ ($M_w = 24$ g/mol) | 0.07 % | 0.09 % |
| $K^+$ ($M_w = 39$ g/mol) | 0.27 % | 0.63 % |
| $Ca^{2+}$ ($M_w = 43, 44$ g/mol) | <0.00 % | <0.00 % |

For the rheological measurements, in this work we prepare aqueous solutions of pure and mixed $\kappa-C$ and $\iota-C$ without any added salt in Millipore water (resistivity 18.2 MΩ.cm). To remove any excess moisture, the $\kappa-C$ and $\iota-C$ powders are oven dried for 2 h at 120°C. We add predetermined amounts of carrageenan in water and stir it at 400 rpm for 2 h using an IKA C-MAG HS7 magnetic stirrer. Throughout the sample preparation time, we maintain the solution temperature at 80°C for better homogenization. The pure solutions of $\kappa-C$ and $\iota-C$ are prepared for concentration range of, $c_p = 1 - 3$ wt.%. For the mixed gels, we carry out the measurements for a total carrageenan concentration of $c_m = 3$ wt.% at various ratios of $\iota-C$ to $\kappa-C$. The concentration 3 wt.% is chosen such that the gelation data of mixed systems can be compared with that of pure systems at equal concentrations as well as at equal individual concentrations of each carrageenan.

In this work, the rheology experiments have been carried out using a Dynamic Hybrid Rheometer 3 by TA instruments. We use a serrated concentric cylinder geometry with 28 mm bob diameter and 1.2 mm gap thickness. The measuring geometry is equipped with a Peltier-Plate temperature control assembly. Each sample is charged into the shear cell at 80°C and brought to 60°C by applying a temperature ramp of $k = -0.5$ °C/min.

Subsequently, to formulate an equilibrium gel structure, we subject the sample to a cooling cycle at a very slow constant ramp rate, $k = -0.05$ °C/min. Followed by the cooling cycle, we keep the sample at 10°C for 1 h and apply a heating cycle at a constant ramp of $k = +0.05$ °C/min. During the application of cooling and heating cycles, we carry out cyclic frequency sweep measurements with angular frequency varying in the range of $\omega = 0.1 - 25$ rad/s at a constant stress amplitude of $\sigma = 0.1$ Pa. To avoid water losses due to evaporation during rheological measurements, the free surface of the sample is always covered with a thin layer of low density silicone oil. In this manuscript, we denote the properties associated with sol-gel and gel-sol transition with subscript $-SG$ and $-GS$, respectively.

## 3. Results:

Pure and mixed systems of $\kappa-C$ and $\iota-C$ are in the sol state at 60°C, which is an equilibrium state. Consequently, any prior shear or thermal history associated with solutions is forgotten. On decreasing temperature below 60°C, eventual hydrogen bond formation between polymeric strands leads to the formation of helices and double helices. With a further decrease in temperature, the double helices aggregate to form a three-dimensional percolated space-spanning network that we represent as a gel. In Figure 3 (a), we illustrate the increase in elastic modulus ($G'$) during cooling of 1.5 wt.% and 3 wt.% of pure $\kappa-C$ and $\iota-C$ and a mixed system with 1.5 wt.% of each $\kappa-C$ and $\iota-C$ at a constant frequency. To avoid clutter, we do not show the evolution of loss moduli ($G''$), which also exhibits an increase with a decrease in temperature. Liu *et al.* (Liu, Huang, et al., 2016) comprehensively report multiple methods of determining the gelation temperature for cooling/ heating cycles carried out at a single frequency. The methods comprise the inception of $G'$ during cooling, the crossover point of $G'$ and $G''$, and the asymptote method for complex viscosity, $\eta^*$. Here, we follow the method of inception of $G'$ during cooling, as followed in the literature of mixed gels of $\kappa-C$ and $\iota-C$.

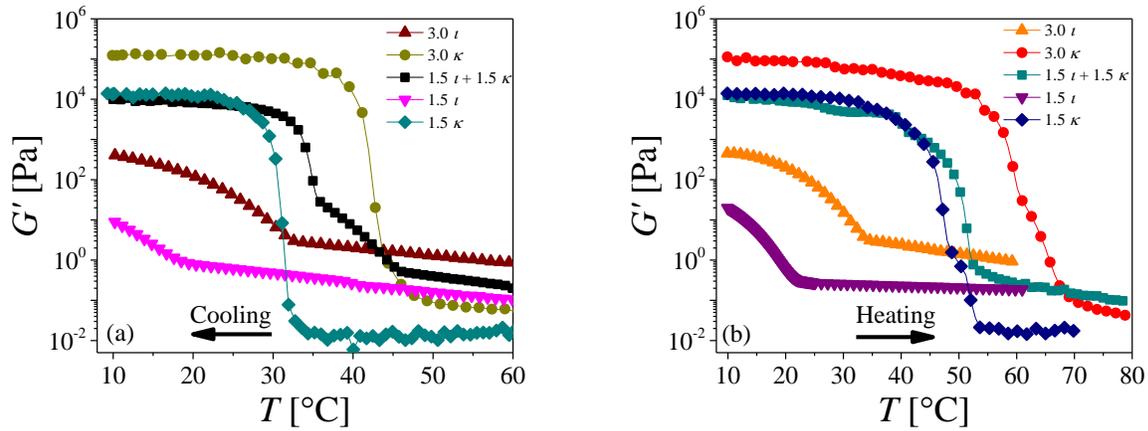

**Figure 3.** The elastic modulus ($G'$) is plotted as a function of temperature at a fixed angular frequency, $\omega = 3.14$ rad/s during (a) cooling and (b) heating at a ramp rate, $k = \pm 0.05$ °C/min for pure and mixed carrageenan gels.

It can be seen in Figure 3 (a) that with a decrease in temperature, $G'$ of pure $\kappa-$C at 1.5 wt.% and 3 wt.% shows a steep one-step increase at roughly 32°C and 47°C respectively. On the other hand, the pure $\iota-$C solutions show a rather gradual increase in $G'$ on cooling with a transition at around 17°C and 31°C for the 1.5 wt.% and 3 wt.% systems, respectively. Very interestingly, the mixed gel shows a two-step increase in $G'$ while cooling. The corresponding transition temperatures associated with the two steps are approximately 45°C and 35°C. According to previous reports, the $\kappa-$C and $\iota-$C components in the mixed gel undergo gelation separately at these temperatures (Bui, Nguyen, Renou, et al., 2019; Du, Brenner, et al., 2016; Geonzon, et al., 2019a; Parker, et al., 1993). Interestingly, these two temperatures are closer to the respective temperatures at which $G'$ starts to increase for pure 3 wt.% $\kappa-$C and $\iota-$C. Our temperature data of two-step change agrees well with some of the previous works (Du, Brenner, et al., 2016; Geonzon, et al., 2019a), where pure ($c_{p,1}$ and $c_{p,1}$ wt.%) and mixed ($c_m$ wt.%) systems are compared at equal total carrageenan concentration ($c_{p,1}=c_{p,2}=c_m$). This also suggests purity level of our samples should be close to the samples investigated in the above studies.

Interestingly, in the sol state at a higher temperature regime, owing to higher charge density and bulkier coils, pure $\iota-$C has a higher modulus compared to pure $\kappa-$C. On the other hand, in the low-temperature regime, pure $\kappa-$C has a higher modulus due to better aggregation of double helices compared to pure $\iota-$C. At the same total carrageenan concentration, the modulus in the mixed system is in between that of pure $\kappa-$C and $\iota-$C in both the high temperature and the low-temperature regimes. After cooling, we maintain the system at 10°C for an hour and apply the heating cycle at $k = +0.05$ °C/min. In Figure 3 (b), we plot the evolution of $G'$ during heating and explore how the network structure dissociates. Interestingly for pure $\kappa-$C systems, the magnitudes of $G'$ start to decrease at a much higher temperature compared to the inception temperature of $G'$ during cooling. On the contrary, $\iota-$C exhibits an excellent thermoreversibility. Intriguingly, contrary to what was observed during cooling, the mixed system also shows only one step decrease in the magnitude of $G'$ as it is heated. Nonetheless, irrespective of the direction of temperature change, the temperatures at which the transition occurs are observed to show an inherent dependence on frequency (not shown). Owing to this, the respective transition points may not be termed as the precise gelation points. To identify the true gelation temperature, we carry out cyclic frequency sweep measurements during the cooling and heating of the pure and mixed systems.

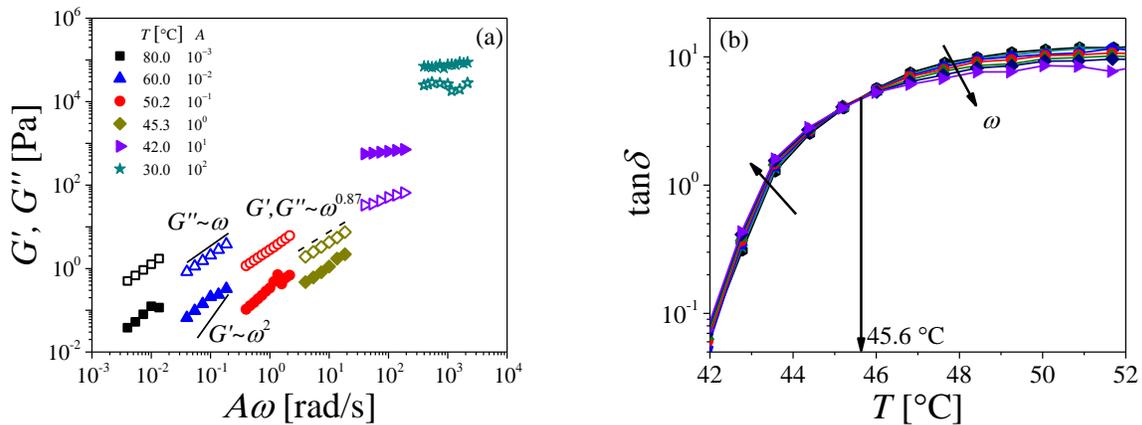

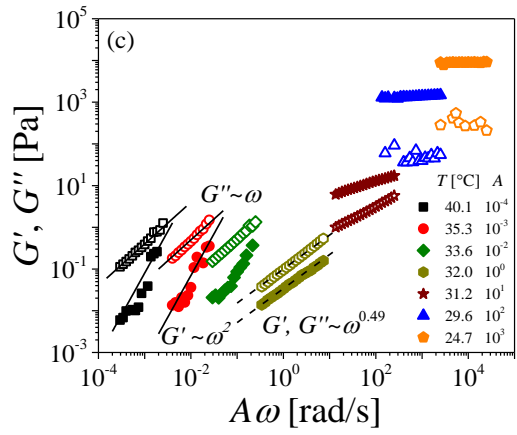
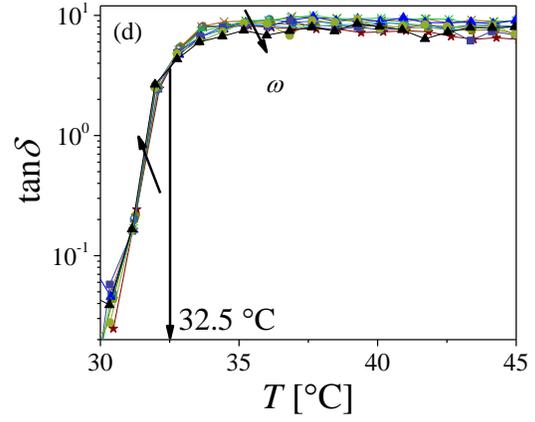
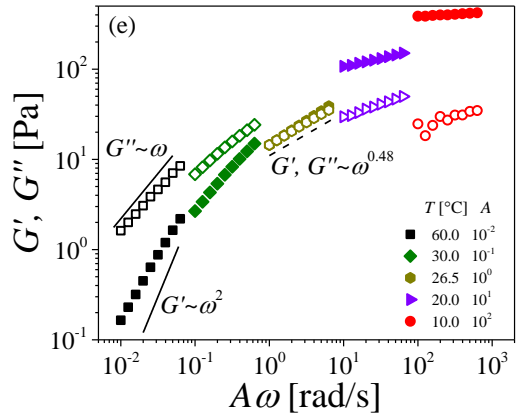
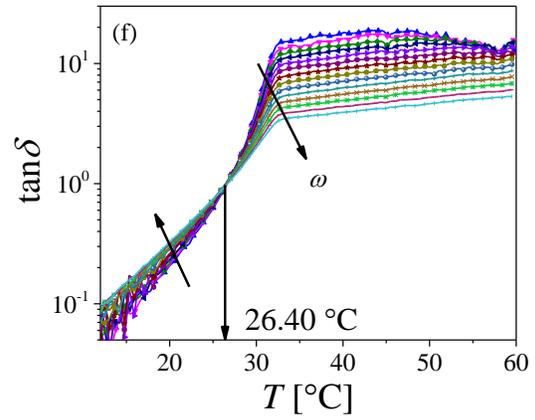
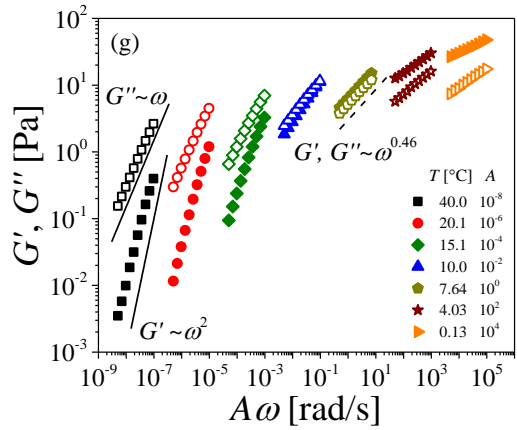
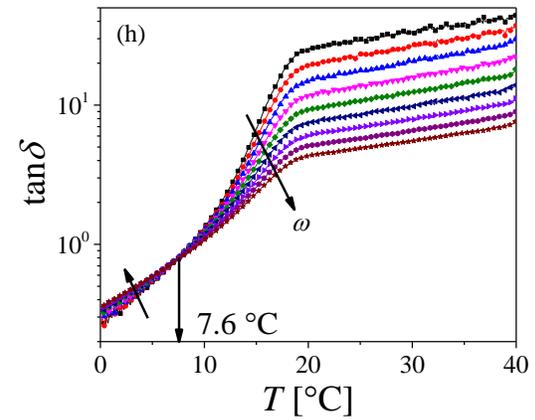

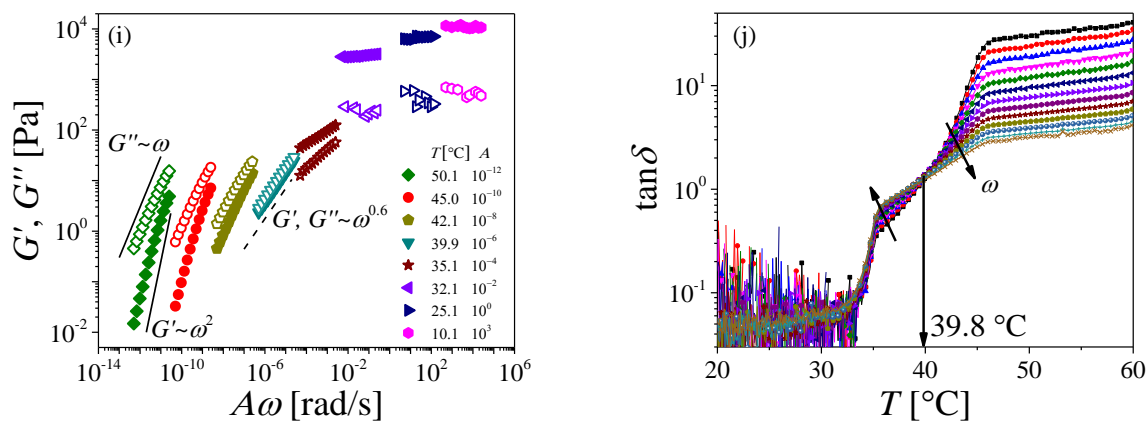

**Figure 4.** The evolution of linear viscoelastic properties has been plotted for the measurements carried out during cooling at a fixed ramp rate, $k = -0.05$ °C/min. (a,c,e,g,i) The dynamic moduli, $G'$ and $G''$ are plotted as a function of $\omega$ at different temperatures. The horizontal axes are shifted with a factor $A$, whose magnitudes are mentioned in the legends. In the low-temperature regime, the systems show a rheological signature of sol state: $G' \sim \omega^2$ and $G'' \sim \omega$, as shown by the solid lines. At the critical gel state, the $G'$ and $G''$ follow the identical power law dependence, as shown by the dashed lines. (b,d,f,h,j) The $\tan\delta$ is plotted as a function of temperature at different applied $\omega$ during the cyclic frequency sweep measurements. The various investigated systems are pure $\kappa-$C of concentrations (a, b) 3 wt.% and (c, d) 1.5 wt.%; pure $\iota-$C of concentrations (e, f) 3 wt.% and (g, h) 1.5 wt.%; and (i, j) mixed carrageenan with 1.5 wt.% of each $\kappa-$C and $\iota-$C.

In Figure 4 we depict the results of cyclic frequency sweep measurements during the cooling of the aqueous solutions of pure and mixed carrageenan. In Figure 4 (a), we plot the evolution of $G'$ and $G''$ as a function of angular frequency, $\omega$ at different temperatures for the pure 3 wt. % $\kappa-$C system. In Figure 4 (b), we plot the corresponding evolution of $\tan\delta$ as a function of decreasing temperature at different $\omega$. At higher temperatures, $G''$ is much higher compared to $G'$, and exhibits signatures of the sol state (terminal region) defined as $G' \sim \omega^2$ and $G'' \sim \omega$ over the explored range of $\omega$. The magnitude of the loss tangent, $\tan\delta = G''/G'$ is higher than unity over this temperature regime. Furthermore, $\tan\delta$ has been observed to increase with an increase in $\omega$, which is also a characteristic feature of the sol state. It can be seen that $\tan\delta$ shows a weak variation with temperatures up to 47°C. With a further decrease in temperature, the dependence of $G'$ and $G''$ on $\omega$ weakens, especially at a higher rate for $G'$ compared to $G''$. At a certain temperature 45.6°C, $G'$ and $G''$ shows an identical power law dependence as: $G', G'' \sim \omega^{0.87}$ as shown in Figure 4 (a). Consequently, $\tan \delta$ becomes independent of frequency at this temperature, as marked with a solid arrow in Figure 4 (b). With the continuing decrease in temperature, both $G'$ and $G''$ grow rapidly as the gel consolidates. Eventually, the dynamic moduli become nearly independent of $\omega$, with $G' \gg G''$. Accordingly, the magnitude of $\tan \delta$ decreases by two orders of magnitude, merely over a temperature variation of 3°C. Such a steep decrease in $\tan \delta$ has also been observed in Figure 3 (d) for the pure 1.5 wt.% $\kappa - C$. On the other hand, the evolution of $\tan \delta$ is more gradual for both the pure $\iota - C$ compared to the pure $\kappa - C$, as shown in Figure 4 (f) and (h). In the case of mixed gel, a two-step decrease is observed in $\tan \delta$ with a change in temperature. Interestingly, the temperature at which the isofrequency curves of $\tan \delta$ merge at a single point, is intermediate of the two-step transition points.

Systems undergoing a sol-gel transition through the physical or chemical crosslinking are known to pass through a unique state called the critical gel state, where the three-dimensional percolated space-spanning network of the gel is at its weakest (H. Henning Winter, et al., 1986; Horst Henning Winter, et al., 1997). According to the seminal work of Winter and Chambon (H. Henning Winter, et al., 1986), the dynamic moduli demonstrate an identical power law dependence on $\omega$ at the critical gel state given by:

$$G' = G'' \cot\left(\frac{n_c \pi}{2}\right) = \frac{\pi S}{2\Gamma(n_c)} \text{cosec}\left(\frac{n_c \pi}{2}\right) \omega^{n_c} \qquad (2)$$

where, $n_c$ is the critical relaxation exponent, and $\Gamma(n_c)$ is the Euler gamma function. The parameter $S$ is known as the gel strength and takes the unit $\text{Pa.s}^{n_c}$. Accordingly, the loss tangent, $\tan \delta = \tan\left(\frac{n_c \pi}{2}\right)$, is independent of $\omega$ at this state. The critical exponent $n_c$ varies between 0 to 1, where, the upper limit of $n_c = 1$ suggests a viscous gel (units of $S$ are that of viscosity), while the lower limit of $n_c = 0$ suggests an elastic gel (units of $S$ are that of the modulus). The gel strength, $S$ is inversely related to $n_c$, suggesting an elastic gel with low $n_c$ will have a higher gel strength (Horst Henning Winter, et al., 1997). In the literature, this is considered to be the rheological signature of the sol-gel transition and has been verified for a variety of colloidal and polymeric systems undergoing gelation transition (Bhattacharyya, et al., 2023; Jatav, et al., 2017; Joshi, et al., 2020; Liu, Bao, & Li, 2016; Liu, et al., 2015; Suman, et al., 2019; Suman & Joshi, 2020; Suman, Mittal, et al., 2020). We measure the true critical sol-gel transition temperature ($T_{c,SG}$) from the data shown in Figure 4 using the Winter-Chambon criteria (H. Henning Winter, et al., 1986). For pure $\kappa-$C of concentration 1.5 wt.% and 3 wt.%, $T_{c,SG}$ = 32.4°C and 45.6°C, respectively, which is very close to the gelation temperature measured from Figure 4 (a). However, for $\iota-$C of concentration 1.5 wt.% and 3 wt.%, $T_{c,SG}$ = 7.6°C and 26.4°C, respectively. These critical sol-gel transition temperatures for $\iota-$C are quite below that obtained from Figure 4 (a), indicating that the method of gelation temperature measurement at the inception point of $G'$ during cooling is not suitable for $\iota-$C. In the case of the mixed gel, we obtain tan $\delta$ to be independent of $\omega$ only at a single temperature in between the two-step change, which we mark as the critical sol-gel transition temperature, $T_{c,SG}$ = 39.8°C as shown in Figure 4 (i) and Figure 4 (j). This suggests that when the total carrageenan concentration is constant at 3 wt.%, such that $c_{p,1}$=$c_{p,2}$=$c_m$, the mixed gel has an intermediate critical sol-gel transition temperature, that is closer to that of pure $\kappa-$C. On the other hand, if we compare the 3 wt.% mixed gel having the equal ratio of $\iota-$C and $\kappa-$C with the pure 1.5 wt.% of each Carrageenan, such that $c_m$=$c_{p,1}$+$c_{p,2}$, the critical sol-gel transition temperature is higher than that of the individual pure components.

In Figure 5, we plot the critical sol-gel transition temperature, $T_{c,SG}$ as a function of concentration for pure and mixed systems of $\kappa-$C and $\iota-$C. For pure systems, we carry out the measurements at 1 wt.%, 1.5wt.%, 2 wt.% and 3 wt.%. As expected, the critical sol-gel transition temperature decreases with a decrease in the concentration of carrageenan, such that the 1 wt.% $\iota-$C does not undergo sol-gel transition within the explored temperature range of 60°C to 0°C. Interestingly, we observe that at any concentration, pure $\iota-$C always has a lower $T_{c,SG}$ compared to the pure $\kappa-$C. This is due to the higher charge density in $\iota-$C, which inhibits the formation of double helices and their aggregation. The slower aggregation process delays the formation of the space-spanning network structure and in turn, decreases the $T_{c,SG}$ in $\iota-$C compared to $\kappa-$C. With respect to the mixed systems, we vary the ratio of $\iota-$C to $\kappa-$C, keeping the total carrageenan concentration constant at 3 wt.%. In this case, $T_{c,SG}$ increases with a decrease in $\iota-$C and an increase in $\kappa-$C fraction in the mixture.

Interestingly, the mixed systems, when compared at the concentration of individual pure components, such that $c_m=c_{p,1}+c_{p,2}$, always show a higher $T_{c,SG}$.

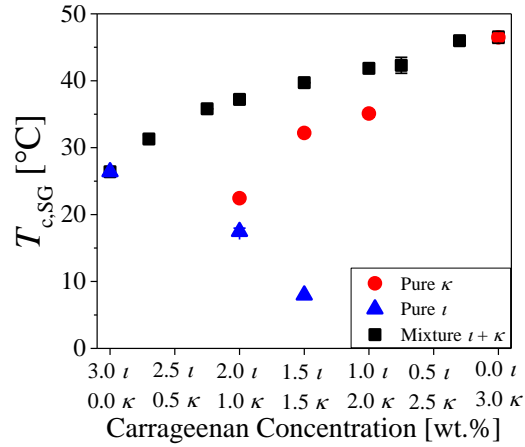

**Figure 5.** The dependence of the critical sol-gel transition temperature $T_{c,SG}$ on concentration is plotted for pure and mixed systems of $\kappa-C$ and $\iota-C$. The cooling measurements are carried out at a ramp rate of $k = -0.05$ °C/min. Please note that for pure $\iota-C$, concentration increases from right to left.

The critical relaxation exponent, $n_{c,SG}$ and gel strength $S_{SG}$, for the critical sol-gel transition state can be obtained from Eq. (2) by analyzing the identical power law dependence of $G'$ and $G''$ on $\omega$, or from the constant value of $\tan\delta$ that is independent of $\omega$, as shown in Figure 4. Furthermore, assuming the critical gel state to be like a branched polymer system with completely screened excluded volume interaction, Muthukumar (Muthukumar, 1989) proposed that fractal dimension, $f_d$ of the percolated network structure at the critical gel state can be computed from the information of critical relaxation exponent, $n_c$ as:,

$$f_d = \frac{5(2n_c - 3)}{2(n_c - 3)} \qquad (3)$$

where, $d$ is dimension of space with value 3 in the present case. Remarkably fractal dimension obtained from this expression corroborates well with that obtained from scattering experiments (Hsu & Yu, 2000; Kara, Arda, & Pekcan, 2018; Matsumoto, Kawai, &

Masuda, 1992; Ponton, Bee, Talbot, & Perzynski, 2005). In Figure 6, we plot the $n_{c,SG}$, $f_{d,SG}$ and $S_{SG}$ as a function of concentration of pure and mixed systems of $\kappa-$C and $\iota-$C. For pure $\kappa-$C, over the concentration regime of 1.5 to 3 wt.%, $n_{c,SG}$ (0.82 − 0.87), $f_{d,SG}$ (1.5 − 1.6), $S_{SG}$ (0.08 − 0.12 Pa. s$^{n_c}$) weakly vary with change in concentration. However, at 1 wt.% concentration, $n_{c,SG}$ suddenly decreases, leading to a more elastic gel with higher values of $f_{d,SG}$ and $S_{SG}$. This may be due to lower $T_{c,SG}$ of the 1 wt.% $\kappa-$C, where more hydrogen bonds are present at the critical gel state, compared to the higher concentrations of $\kappa-$C. In case of pure $\iota-$C in the concentration regime of 1.5 wt.% to 3 wt.%, $n_{c,SG}$ (≈0.45) and $f_{d,SG}$ (≈2.06) are independent of concentration, while $S_{SG}$ (8.5 − 5 Pa. s$^{n_c}$) varies weakly with decrease in concentration. While $\iota-$C at 1% concentration does not show gel transition in the explored temperature range, we can approximately estimate the behavior with concentration change from the data of 1 wt.% $\kappa-$C. We predict that if at all 1 wt.% $\iota-$C gels at subzero temperature without ice formation, the critical gel will have a low $n_{c,SG}$, with high $f_{d,SG}$ and $S_{SG}$. Interestingly, at any concentration, where the system undergoes sol-gel transition, the pure $\iota-$C has a lower $n_{c,SG}$, and higher $f_{d,SG}$ and $S_{SG}$ compared to the pure $\kappa-$C. This suggests that $\iota-$C, having lower $T_{c,SG}$, forms a more elastic, denser, and stronger network compared to $\kappa-$C at the critical sol-gel transition state. For the mixed systems of $\iota-$C and $\kappa-$C, while $n_{c,SG}$ monotonically increases, $f_{d,SG}$ and $S_{SG}$ monotonically decreases with decrease in $\iota-$C and increase in $\kappa-$C fraction in the mixture. Thus, the behavior in mixed gel is intermediate of that of the pure $\kappa-$C and $\iota-$C.

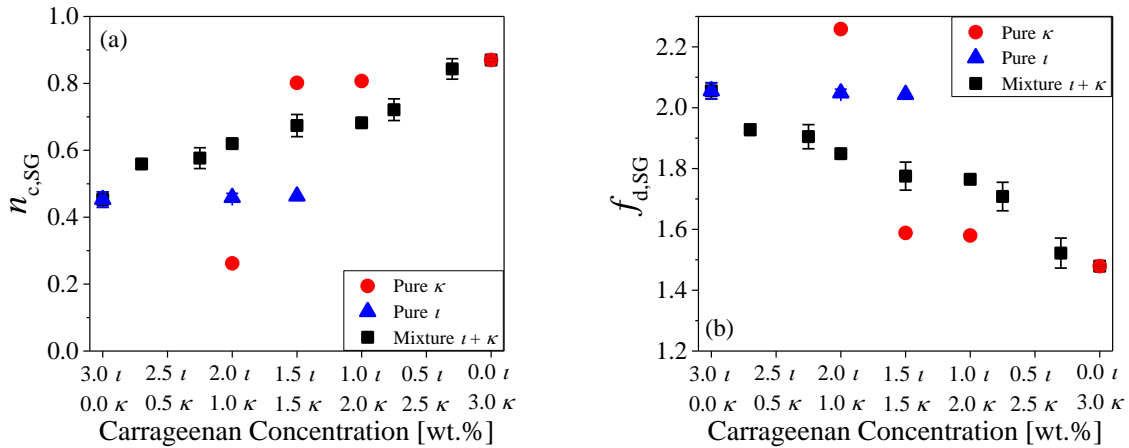

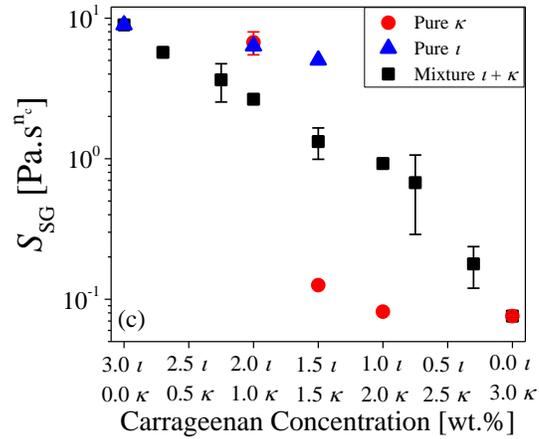

**Figure 6.** (a) The critical relaxation exponent, $n_{c,SG}$, (b) the fractal dimension, $f_{d,SG}$, and (c) the gel strength, $S_{SG}$ associated with the sol-gel transition is plotted as a function of concentration for pure and mixed systems of $\kappa-$C and $\iota-$C. The cooling measurements are carried out at a ramp rate of $k = -0.05$ °C/min. Please note that for pure $\iota-$C, concentration increases from right to left.

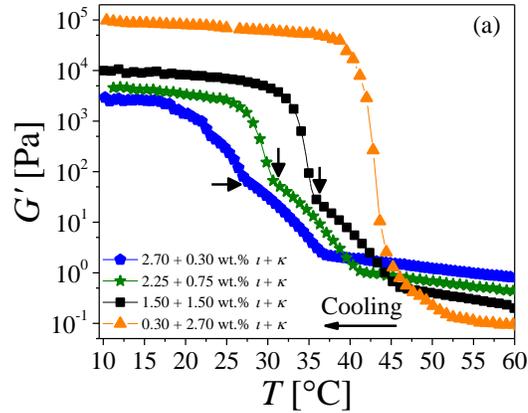

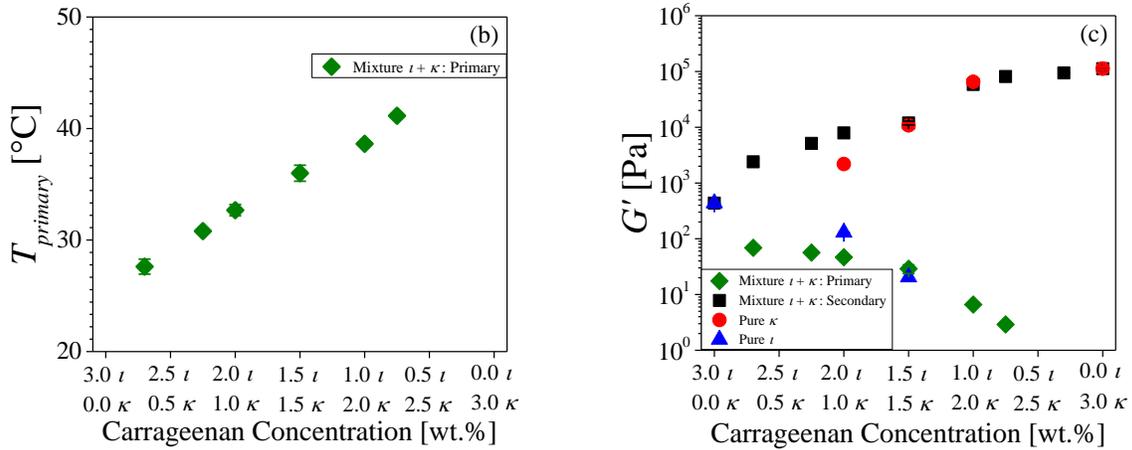

**Figure 7.** (a) The temperature dependence of elastic modulus at a constant $\omega = 3.14$ rad/s during cooling for mixed carrageenan gels. The mixed carrageenan gels exhibit two-step change in moduli during cooling. The plateau moduli at higher temperatures are denoted as primary, and the plateau moduli at lower temperatures are denoted as secondary. The point corresponding to the primary moduli is shown with a solid arrow. (b) The temperature associated with primary modulus is plotted as a function of concentration during cooling for mixed systems of $\kappa-$C and $\iota-$C. (c) The primary and secondary elastic moduli are plotted as a function of concentration during cooling for pure and mixed systems of $\kappa-$C and $\iota-$C. All the cooling measurements are carried out at a ramp rate of $k = -0.05$ °C/min. Please note that in Figure 7 (b) and (c), for pure $\iota-$C, concentration increases from right to left.

In Figure 7 (a), we plot the evolution of $G'$ as a function of temperature during cooling for mixed gels for different ratios of $\iota-$C to $\kappa-$C. As the mixed gels undergo sol-gel transition during cooling, a two-step change is always observed in $G'$ and $\tan\delta$, when there is at least 25 % of $\iota-$C in the mixture. As discussed before, the values of temperature corresponding to the two-step change have been considered as the individual gelation temperatures in the literature (Bui, Nguyen, Renou, et al., 2019). On the other hand, according to our observations that are based on the Winter-Chambon criteria, the true gelation temperature in the mixed systems lies in between the values of temperature associated with two-step change in the modulus. According to Winter and Mours (Horst Henning Winter, et al., 1997), if individual phases of mixed system undergo gelation at two different temperatures, the critical gel point may appear in between the two. This leads to a possibility that, in the present case, the mixed systems are microphase separated in nature, and the critical sol-gel transition temperature observed in Figure 4 is a result of gelation in the induvial phases. However, as observed in Figure 7 (a), the temperature related to the left-hand side step in the two-step transition increases with a decrease in $\iota-$C fraction in the mixed system. In Figure 7 (b) we plot the decrease in temperature associated with primary moduli with decrease in $\iota-$C fraction in the mixture. Since, in our measurements, the first step change in the modulus closely relates to the true gelation temperature of $\kappa-$C, it is impossible that decreasing fraction of $\iota-$C in the mixed system will individually undergo gelation at a higher temperature. This analysis, therefore, suggests that it is highly unlikely that the critical sol-gel transition temperature gets affected by microphase separation in the mixed system.

Our experimental data on the frequency dependence of dynamic moduli shows that the sol-gel transition takes place between the two step-change. Hence, we denote the moduli at the second step change as the primary moduli of the gel. We consider the moduli in the low temperature regime as the secondary moduli of gel. In Figure 7 (c), we plot the primary and secondary moduli of the mixed gels as a function of the concentration of $\iota-$C and $\kappa-$C. As the second step shifts to a higher temperature with a decrease in $\iota-$C fraction in the mixture, the magnitude of primary modulus also decreases. On the other hand, the secondary modulus increases strongly with a decrease in $\iota-$C fraction and an increase in $\kappa-$C fraction. We also plot the modulus of the pure systems of $\kappa-$C and $\iota-$C at 10°C in Figure 7 (c), except for the 1.5wt.% $\iota-$C, where $T_{c,SG}$ is below 10°C and we take the value of modulus at 0°C. Strikingly, for the same concentration, the modulus of the $\kappa-$C system is three orders of magnitude higher compared to that of $\iota-$C. This suggests that, while strands of $\iota-$C with higher charge density form a weak network, the double helices in $\kappa-$C with lower charge density appear to be forming a stronger network. Accordingly, $\iota-$C gels are soft, and $\kappa-$C gels are rigid with a denser network structure.

An interesting observation is the replacement of 0.1 fraction of $\iota-$C by $\kappa-$C and vice-verse in pure carrageenan concentration of 3 wt.%. When the 0.1 fraction of $\iota-$C is replaced by $\kappa-$C, we observe a significant increase in $T_{c,SG}$, $n_{c,SG}$ and secondary moduli, while $S_{SG}$ decreases. On the other hand, when 0.1 fraction of $\kappa-$C is replaced by $\iota-$C, the $T_{c,SG}$, $n_{c,SG}$, $S_{SG}$ and secondary moduli remain unchanged. Interestingly, the former system also shows two-step transition like the other mixed gels, while the latter shows a steep increase in $G'$ like the pure $\kappa-C$ system. This suggests that the effect of $\kappa-C$ on $\iota-C$ is more pronounced than effect of $\iota-C$ on $\kappa-C$, indicating the two carrageenan do not affect each other in a similar fashion in the mixed system.

Subsequent to cooling, the samples are stored at 10°C for an hour and subjected to cyclic frequency sweep during the application of a heating cycle at a ramp rate of $k = +0.05$ °C/min. In Figure 8, we show the evolution of $G'$ and $G''$ with $\omega$ at different temperatures for the pure and mixed carrageenan systems. For all the systems, $G'$ is observed to be higher than $G''$ in the lower temperature regime with a weak dependence on $\omega$, suggesting a gel state. With an increase in temperature, the magnitude of $G'$ and $G''$ decreases, while their dependence on $\omega$ increases. At a certain temperature, both $G'$ and $G''$ exhibit identical power law dependence, suggesting the critical gel-sol transition state. With further increase in temperature, we observe $G'' > G'$ and follow the signature of sol state as $G' \sim \omega^2$ and $G'' \sim \omega$. To be noted here, since the isofrequency $\tan\delta$ curves merge at below 10°C for pure 1.5wt.% $\iota-C$, as shown in Figure 4 (h), it is cooled and stored at 0°C instead of 10°C before applying the heating cycle. Figure 8 is qualitatively similar to Figure 4 for dynamic moduli obtained in a cooling step. Since the variation of $\tan\delta$ is also equivalent, we do not show the same. Previous studies on poly(vinyl alcohol) (Bhattacharyya, et al., 2023; Joshi, et al., 2020) and $\kappa-C$ (Laureano-López, Pérez-López, Espinosa-Solares, & del Carmen Núñez-Santiago, 2022) has verified applicability of Eq. (2) for temperature controlled gel-sol transition. Accordingly, we compute critical gel-sol transition temperature, $T_{c,GS}$, and the corresponding critical relaxation exponent, $n_{c,GS}$ and gel strength, $S_{GS}$. We also obtain the fractal dimension associated with the critical gel-sol transition state, $f_{d,GS}$ according to Eq. (3), proposed by Muthukumar (Muthukumar, 1989).

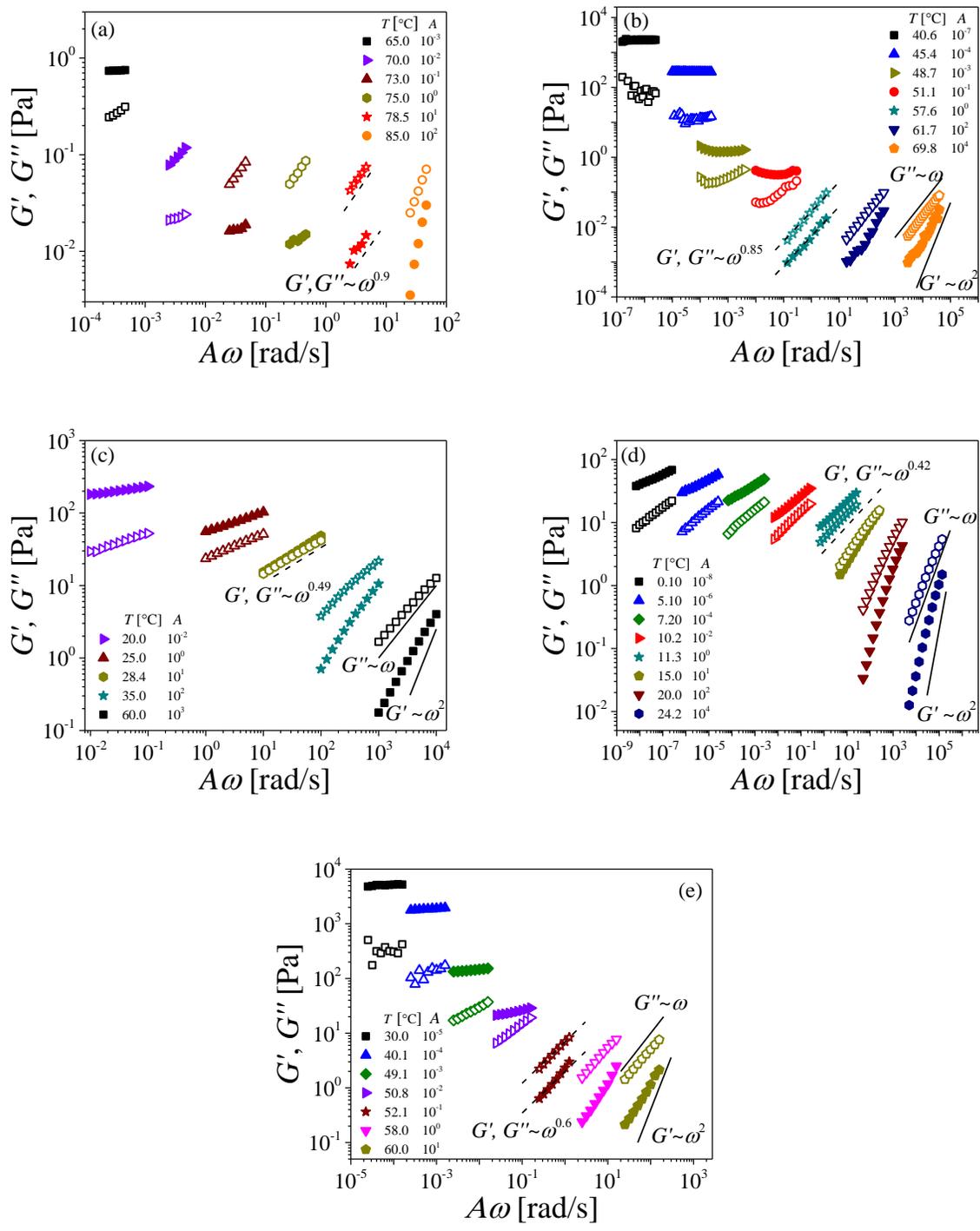

**Figure 8.** The $G'$ and $G''$ are plotted as a function of $\omega$ at different temperatures. The horizontal axes are shifted with a factor $A$, whose values are mentioned in the legends. At the critical gel state, the $G'$ and $G''$ follow the identical power law dependence, as shown by the dashed lines. In the high temperature regime, the systems show rheological signatures of the sol state: $G' \sim \omega^2$ and $G'' \sim \omega$, as represented by the solid lines. The various investigated systems are pure $\kappa-$C of concentration (a) 3 wt.% and (b) 1.5 wt.%; pure $\iota-$C of concentration (c) 3 wt.% and (d) 1.5 wt.%; and (e) mixed carrageenan with 1.5 wt.% of each $\kappa-$C and $\iota-$C. All the measurements are carried out during heating at a fixed ramp rate, $k = +0.05$ °C/min.

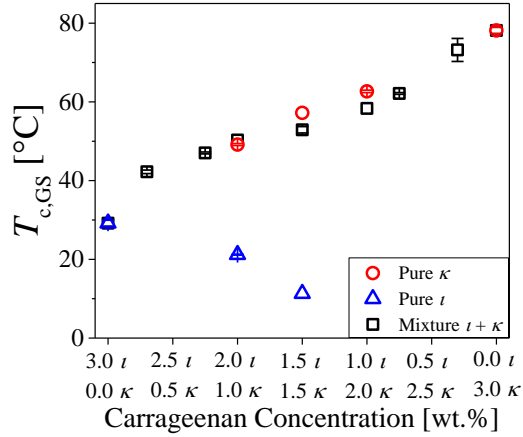

**Figure 9.** The dependence of critical gel-sol transition temperature $T_{c,GS}$ is plotted as a function of concentration for pure and mixed systems of $\kappa-$C and $\iota-$C. The heating is carried out at a ramp rate of $k = +0.05$ °C/min. Please note that for pure $\iota-$C, concentration increases from right to left.

In Figure 9 we plot $T_{c,GS}$ as a function of concentration for the pure and mixed systems of $\kappa-$C and $\iota-$C. Both the pure systems show an increase in $T_{c,GS}$ with an increase in concentration. Interestingly, the pure $\iota-$C shows similar temperature for the gel-sol transition as observed for the sol-gel transition ($T_{c,GS} \approx T_{c,SG}$) which could be because of a loose aggregation of double helices in the same (Geonzon, et al., 2023). On the other hand, the pure $\kappa-$C shows a strong thermal hysteresis wherein the temperature for the gel-sol transition is significantly greater than that of the sol-gel transition ($T_{c,GS} \gg T_{c,SG}$), which we feel could be due to the stronger aggregation in $\kappa-$C double helices that requires higher temperatures to dissociate. The mixed gels of $\kappa-$C and $\iota-$C with a total carrageenan concentration of 3 wt.%, show an increase in $T_{c,GS}$ with a decrease in $\iota-$C fraction and an increase in $\kappa-$C fraction in the mixture, as also observed for $T_{c,SG}$. In Figure 10, we plot $n_{c,GS}$, $f_{d,GS}$ and $S_{GS}$ as a function of concentration for the pure and mixed systems of $\kappa-$C and $\iota-$C. Interestingly, for all the explored concentrations of pure $\kappa-$C and pure $\iota-$C, $n_{c,GS}$, $f_{d,GS}$ and $S_{GS}$ depend weakly on the concentration. As discussed before, the 1 wt.% pure $\kappa-$C system demonstrated a much lower $T_{c,SG}$ and higher $n_{c,SG}$ compared to other higher explored concentrations of pure $\kappa-$C. However, during gel-sol transition, the critical gel properties of 1 wt.% pure $\kappa-$C system remarkably also vary weakly with concentration. This suggests that once a network is formed in $\kappa-$C systems, the aggregation is so strong that higher temperatures are required to dissociate the same. Furthermore, owing to lower values of $T_{c,GS}$, the network structure of pure $\iota-$C at critical gel-sol transition state is denser compared to $\kappa-$C. This means, unlike $\iota-$C, $\kappa-$C systems remain in the gel state while heating, for those temperatures that are in sol state while cooling. The mixed systems of $\kappa-$C and $\iota-$C also show intermediate behavior with respect to 3 wt.% pure $\kappa-$C and $\iota-$C, as observed for other mixed gel properties.

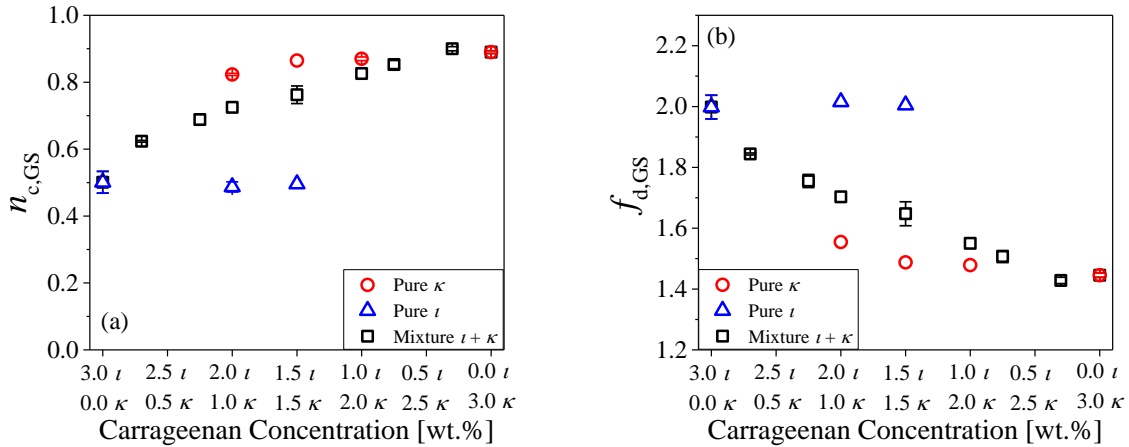

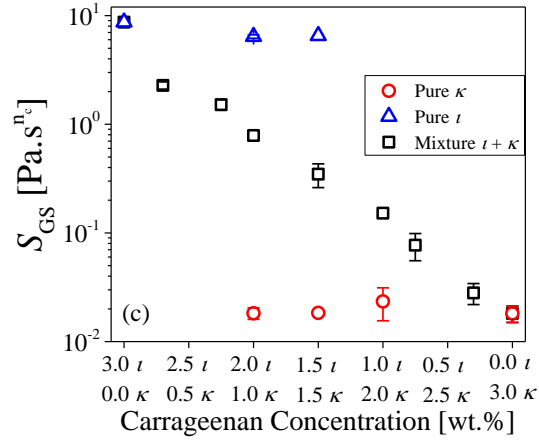

**Figure 10.** (a) The critical relaxation exponent, $n_{c,GS}$ (b) the fractal dimension, $f_{d,GS}$ and (c) the gel strength, $S_{GS}$ associated with gel-sol transition is plotted as a function of concentration for pure and mixed systems of $\kappa-$C and $\iota-$C. The heating is carried out at a ramp rate of $k = +0.05$ °C/min. Please note that for pure $\iota-$C, concentration increases from right to left.

Next, we compare the mixed systems with pure systems of $\kappa-$C and $\iota-$C at concentrations equal to that of the individual component in the mixed system. This means that the comparison is made for $c_m$ wt.% of mixed systems with $c_{p,1}$ wt.% of pure $\kappa-$C and $c_{p,2}$ wt.% of pure $\iota-$C, such that $c_m=c_{p,1}+c_{p,2}$. Under such circumstances, we analyze three types of mixed systems: Type I: $c_{p,1}=1$ wt.% and $c_{p,2}=2$ wt.%; Type II: $c_{p,1}=1.5$ wt.% and $c_{p,2}=1.5$ wt.%; Type III: $c_{p,1}=2$ wt.% and $c_{p,2}=1$ wt.%. Interestingly, in the case of sol-gel transition, mixed systems with $c_m$ wt.% has higher $T_{c,SG}$ compared to pure systems of $\kappa-$C and $\iota-$C for all the types. This indicates an associative interaction of the two carrageenan, which aids in the formation of a three-dimensional network structure at a higher temperature. For the other properties at critical sol-gel transition state, two distinct behavior can be observed. Firstly, for the Type I system, where $\kappa-$C fraction is less than half, $n_{c,SG}$ is much higher, but $f_{d,SG}$ and $S_{SG}$ is lower than the individual pure systems. The low temperature secondary modulus is also higher than the individual $\kappa-$C and $\iota-$C. Secondly, for the Type II and Type III systems, where $\kappa-$C fraction is more than half, $n_{c,SG}$, $f_{d,SG}$ and $S_{SG}$ are intermediate of the pure $\kappa-C$ and $\iota-C$ systems. For such conditions, the low temperature secondary moduli are close to that of pure $\kappa-C$ with concentration $c_{p,1}$ wt.%. This suggests a strong influence of $\kappa-C$ aggregates in the gelation behavior of mixed systems. Since the Type I system contains more $\iota-C$, addition of $\kappa-C$ significantly increases both $T_{c,SG}$ and the low temperature secondary modulus, and the more viscous nature (higher $n_{c,SG}$ and lower $S_{SG}$) of mixed system at critical gel state is only due to higher $T_{c,SG}$. However, when the mixed systems have at least half of $\kappa-C$, the addition of $\iota-C$ moderately increases the gelation temperature but does not change the secondary moduli. This advocates the gel consolidation in mixed gels to be primarily driven by aggregation of double helices of $\kappa-C$. Interestingly, in the case of gel-sol transition, $T_{c,GS}$ of mixed systems are always observed to be equal to that of $\kappa-C$, as also observed by Piculell *et al.* (Piculell, et al., 1989). The other properties associated with the critical gel-sol transition state, $n_{c,GS}$, $f_{d,GS}$ and $S_{GS}$ behaves intermediately. However, variation of $n_{c,GS}$ and $f_{d,GS}$ is such that they are closer to pure $\kappa-C$. This could be because of presence of strong aggregates of $\kappa-C$, that survive during the heating stage, such that their dissociation dictates the gel-sol transition. However, owing to the presence of strands of both the carrageenan, the strength of the critical gel state remains intermediate. This further confirms the gel-consolidation of mixed systems in the low temperature regime to be controlled by $\kappa-C$ aggregation.

## 4. Discussions:

To comprehend the nature of microstructure in the mixed gel systems of $\kappa-C$ and $\iota-C$, it is vital to understand the nature of gel formation in the pure systems. Upon cooling, the hydrogen bonds that get formed between the intrachain and interchain hydroxyl groups of carrageenan molecules lead to a transition from random coil to double helices. The consequent aggregate formation eventually produces the three-dimensional percolated space-spanning network structure. While $\kappa-C$ systems are reported to form a rigid gel with closely aggregated double helices, the $\iota-C$ systems form softer gels with loose aggregates. The addition of monovalent or divalent cations to this aqueous system neutralizes the charge associated with the anionic sulphate ester groups. Accordingly, the addition of cations strongly favors the aggregation process of the double helices, leading to a more consolidated gel network structure (Liu, et al., 2015; Liu & Li, 2016; Nguyen, Nicolai, Benyahia, & Chassenieux, 2014). For any system to undergo a sol-gel transition, a three-dimensional network must get formed, whether be it through the loose or strong aggregation of double helices. In the literature, $\iota-$C systems are reported to show the inception of $G'$ (at a single frequency) at a higher temperature compared to $\kappa-$C in presence of added salts of potassium and calcium (Bui, Nguyen, et al., 2019a; Bui, Nguyen, Renou, et al., 2019; Geonzon, et al., 2019a). However, in the present work with very less concentration of $K^+$, $Mg^{2+}$ and $Ca^{2+}$, we observe inception of $G'$ at a later temperature in $\iota-$C, compared to $\kappa-$C. Subsequently, we also show that the method of estimation of gelation temperature that employs inception of $G'$ is misleading as the scale invariant critical sol-gel transition occurs at a much later temperature in $\iota-$C. Interestingly, our critical sol-gel transition temperature data also matches well with previous literature on $\kappa-$C (Liu, Huang, et al., 2016) and $\iota-$C (Hossain, et al., 2001).

Owing to the higher charge density in pure $\iota-$C, the formation of double helices is less favorable in the same compared to $\kappa-$C. Furthermore, as the strands of $\iota-$C repel each other more compared to $\kappa-$C, formation of a three-dimensional network in the same is sluggish. Accordingly, it seems obvious that the critical sol-gel transition state of $\iota-$C is achieved at a lower temperature compared to $\kappa-$C. The sluggish aggregation process in $\iota-$C leads to a more gradual increase in the modulus compared to $\kappa-$C, thus eventually resulting in a softer gel. Furthermore, as shown in Figure 4, the greater value of $G'$ of $\iota-$C systems, compared to $\kappa-$C systems at higher temperatures may originate from its more opened-up structure attributed to the higher repulsion between the polymeric strands. However, since the process of aggregation and gel formation in carrageenan is more complex in the presence of monovalent or divalent cations, it may be possible that under certain conditions, $\iota-$C sol-gels at a higher temperature. Accordingly, the $\iota-$C systems in the presence of salt need to be more judiciously investigated through the Winter-Chambon criteria (H. Henning Winter, et al., 1986). As discussed previously, owing to the lower critical sol-gel transition temperature, $\iota-$C possesses a more elastic, closed, and stronger network at the critical gel state than $\kappa-$C. However, beyond the critical sol-gel transition state, as the double helices of $\kappa-$C aggregate more closely compared to $\iota-C$ due to the lower charge density, the intrinsic behavior of $\iota-C$ and $\kappa-C$ gels are reflected in the low-temperature regime, where $\kappa-C$ gels are rigid and $\iota-C$ network gels are soft.

The distinction between microphase separation and independent interpenetrated network formation has been made primarily through rheological measurements. As discussed before, many researchers suggest moduli of mixed gels with independent interpenetrated networks to be a summation of individual moduli of pure components. However, according to Ridout *et al.* (Ridout, et al., 1996), the modulus of mixed gel with an independent interpenetrated network must be dominated by the stiffer network, which is $\kappa-C$ in the present case. This matches our results very well, as we observe the secondary moduli of pure $\kappa-C$ and mixed systems to be similar at the equal concentration of individual components. Despite that, we believe when moduli of pure and mixed systems strongly depend on temperature, as observed by us and others (Bui, Nguyen, Renou, et al., 2019; Geonzon, et al., 2019a), utilizing them to distinguish the microphase separation and independent interpenetration may result in misleading conclusions. Accordingly, Eq. (1) should only be used for those situations where pure and mixed gel moduli are independent of the state cooling cycle for a sufficient span of temperatures. Nonetheless, the turbidity and microscopic images by Bui *et al.* (Bui, Nguyen, et al., 2019a; Bui, Nguyen, Renou, et al., 2019) conclusively establish that microphase separation is not possible in mixed gels of $\kappa-C$ and $\iota-C$.

The independent interpenetrated network formation is only possible when the concentrations of both components are such that they can form separate networks. Otherwise, one component will be present in the other only as individual helices, non-participating in a continuous network structure. In this regard, the effect of the replacement of small amounts of $\kappa-C$ in $\iota-C$ and vice-verse should be ideally the same. On replacement of 0.1 fraction of $\iota-C$ by $\kappa-C$, the critical sol-gel transition temperature increases by 20%, while the secondary moduli in the low-temperature regime exhibit a 400% increase. However, on replacement of 0.1 fraction of $\kappa-C$ by $\iota-C$, the system does not demonstrate any significant changes in properties, neither at the critical gel state nor in the low-temperature regime. Furthermore, a prominent two step change is observed in the former case, while it is absent in the latter. This suggests that not only $\kappa-C$ affect $\iota-C$ more significantly, it enhances the ability of aggregation of the $\iota-C$ strands in such a way that the critical gel state is observed at a higher temperature, and the gel is much denser in the low-temperature regime. This is contradictory to independent interpenetrated network formation, and rather implies a network where $\iota-C$ and $\kappa-C$ coexist such that a stronger network of $\kappa-C$ dominates the properties. However, a modified interpenetrated network may form, as suggested by Bui *et al.* (Bui, Nguyen, Renou, et al., 2019).

The possibilities of coexisting network structure formation in mixed gels are either through mixed helices formation or through coaggregation of the double helices. In the literature (Parker, et al., 1993), the possibility of the formation of mixed helices has always been rejected due to the two-step changes observed in the mixed systems. However, we conclusively show that the two-step change is not directly related to the critical sol-gel transition point. Furthermore, inconsistency in the comparison of the gelation transition state of the mixed gel with pure components suggests that the match of the two-step change of the mixed system with the corresponding pure system is only accidental. Given that the formation of mixed helices of $\iota-C$ / $\kappa-C$ is electrostatically more favorable than pure $\iota-C$ double helices (Parker, et al., 1993), we believe such network structure is equally probable in mixed systems.

Our critical sol-gel transition temperature, $T_{c,SG}$ data in Figure 5 suggests that when the two carrageenan moieties are mixed (such that, $c_m=c_{p,1}+c_{p,2}$), they undergo gelation at a higher temperature, compared to the individual concentration of pure components. Furthermore, the gel strength of the mixed system at the critical sol-gel transition state is more than the two when it contains more $\iota-C$, and is intermediate of the two when it has more $\kappa-C$. Such formation of a denser network at a higher temperature compared to the individual concentration of pure components is only possible when the strands of $\kappa-C$ and $\iota-C$ are interacting with each other. When the long chains of $\iota-C$ and $\kappa-C$ interact with each other, there will be a greater number of double helices, even at a higher temperature. Both formations of mixed helices and coaggregation of double helices will lead to the critical gel state at an earlier temperature, compared to that of the pure components.

If $\kappa-C$ and $\iota-C$ strands coexist, their difference in aggregation pattern due to different charge densities is bound to be prominent in mixed gel behavior. According to our results, when small amounts of $\kappa-C$ in pure systems are replaced by $\iota-C$ in a controlled fashion, the presence of $\iota-C$ decreases the $T_{c,GS}$ and the secondary modulus only by 5% and 16%, respectively. Interestingly, the properties of mixed systems are strongly dominated by $\kappa-C$, when at least half of the mixed systems consist of $\kappa-C$. With a further decrease in the fraction of $\kappa-C$ in the mixed system, the system behavior deviates from $\kappa-C$, but does not replicate the pure $\iota-C$ behavior. Furthermore, the thermoreversibility of $\iota-C$ in a mixed system is also altered considerably, such that the gel-sol transition is driven by $\kappa-C$. This is due to the lower charge density in $\kappa-C$, which promotes stronger aggregation of $\iota-C$ in the mixed gel network and delays the network structure dissociation with an increase in temperature. On the other hand, higher charge density in $\iota-C$ hinders the close presence of chains, resulting in a gradual aggregation and, thus, the formation of gels. Since the primary modulus does not correspond to gelation temperature, a logical explanation of its occurrence essentially arises from this inhibition in the aggregation process due to the presence of $\iota-C$. As more intrachain and interchain hydrogen bonds get formed with a decrease in temperature, electrostatically favorable aggregation of $\kappa-C$ strands starts to dominate the gel consolidation. Accordingly, the modulus starts to increase with a decrease in temperature as the $\kappa-C$ strands aggregate with each other or with $\iota-C$ strands. The mixed helices formation or coaggregation of double helices significantly increases the moduli observed in the low-temperature regime for pure $\iota-C$, as reported in this work as well as others (Bui, Nguyen, Renou, et al., 2019; Du, Brenner, et al., 2016; Geonzon, et al., 2019a; Piculell, et al., 1992).

## 5. Conclusions:

The present work comprehensively investigates the rheological signatures of sol-gel and gel-sol transition in pure and mixed systems of $\kappa-C$ and $\iota-C$ by probing the systems with cyclic frequency sweep during cooling and heating. We compare the $T_{c,SG}$ observed for cyclic frequency sweep with the measurement of gelation transition temperature through the method of inception of $G'$ point at a single frequency. Both the methods yield comparable results for $\kappa-$C, but in the case of $\iota-$C the true gelation temperature $T_{c,SG}$ is much lower than the temperature at which $G'$ starts to increase. On the other hand, while the distinct two-step increase in $G'$ of mixed systems is proposed to be associated with the gelation temperature of each component, we observe the scale-invariant $T_{c,SG}$ to lie between the two steps. Furthermore, the mixed gels demonstrate a higher $T_{c,SG}$ compared to the pure individual systems, suggesting the coexistence of $\kappa-$C and $\iota-$C in mixed systems. We observe that when at least half of the mixed system constitutes $\kappa-$C, the low-temperature moduli are closer to $\kappa-$C, and the presence of $\iota-$C does not obstruct the aggregation of double helices of $\kappa-$C. On the other hand, when on 0.1 fraction of $\iota-$C is replaced by $\kappa-$C in pure 3 wt.% $\iota-$C system, low-temperature moduli increase by four times. Furthermore, the presence of $\kappa-$C in $\iota-$C also strongly impairs the thermoreversibility behavior of $\iota-$C. This suggests that the effect of $\kappa-$C on $\iota-$C and vice-verse is not similar, which is only possible when they have an interacting network structure dominated by $\kappa-$C. These rheological measurements strongly advocate the formation of the associative network in mixed gels of $\kappa-$C and $\iota-$C, wherein the lower charge density of the former governs the aggregation and dissociation of overall aggregates.

6. **Acknowledgements:**

We would like to acknowledge financial support from Science and Engineering Research Board, Government of India.

7. **Declaration of Interests:** The authors have no conflicts of interest to disclose.

8. **Data Availability Statement:** The experimental data and information regarding the Carrageenan samples will be disclosed upon a reasonable request.

## 9. List of symbols

| | |
|---|---|
| $c_m$ | Concentration of mixed system |
| $c_p$ | Concentration of pure system |
| $d$ | Dimension of space |
| $f_d$ | Fractal dimension |
| $f_{d,SG}$ | Fractal dimension associated to sol - gel transition |
| $f_{d,GS}$ | Fractal dimension associated to gel - sol transition |
| $G_m$ | Elastic modulus of mixed system |
| $G_p$ | Elastic modulus of pure system |
| $G'$ | Elastic modulus [Pa] |
| $G''$ | Viscous modulus [Pa] |
| $k$ | Cooling/ Heating ramp rate [°C/$min$] |
| $M_w$ | Molecular weight [g/mol] |
| $n_c$ | Critical relaxation exponent [-] |
| $n_{c,SG}$ | Critical relaxation exponent associated to sol - gel transition [-] |
| $n_{c,GS}$ | Critical relaxation exponent associated to gel - sol transition [-] |
| $S$ | Gel strength [Pa.$s^{n_c}$] |
| $S_{SG}$ | Gel strength associated to sol - gel transition |
| $S_{GS}$ | Gel strength associated to gel - sol transition |
| $T$ | Temperature [°C] |
| $T_{c,SG}$ | Critical sol - gel transition temperature [°C] |
| $T_{c,GS}$ | Critical gel - sol transition temperature [°C] |

**Greek Symbols:**

| | |
|---|---|
| $\eta^*$ | Complex viscosity [Pa.s] |
| $[\eta]$ | Intrinsic viscosity [ml/gm] |
| $\varphi$ | Phase volume fraction of each component in mixed gel |
| $\chi$ | Exponent associated with Polymer Blending Law |
| $\omega$ | Angular frequency [rad/s] |